\documentstyle[aps,twocolumn,epsf]{revtex}
\begin{document}
\draft
\title{Black holes: new horizons} 
\author{Sean A. Hayward}
\address{Asia Pacific Center for Theoretical Physics,\\
The Korea Foundation for Advanced Study Building 7th Floor,
Yoksam-dong 678-39, Kangnam-gu, Seoul 135-081, Korea\\
{\tt hayward@mail.apctp.org}}
\date{Revised 26th October 2000}
\maketitle

\begin{abstract}
This is a review of current black-hole theory,
concentrating on local, dynamical aspects.
(Expanded version of the brief talk given 
to open the Generalized Horizons session 
of the Ninth Marcel Grossmann Meeting, Rome, July 2000).
\end{abstract}

\section{Brief history}

The eve of a new millenium provides a particularly opportune occasion
to review the status of any enterprise and its foreseeable future.
The study of black holes is relatively young by scientific timescales;
although Schwarzschild\cite{Sch} found the first such solution in 1916,
almost immediately after Einstein formulated General Relativity \cite{Ein},
its global structure was described only in 1960 
with the paper of Kruskal\cite{Kru}, reportedly written by Wheeler,
who is acknowledged as coining the term black hole in the late 1960s\cite{Whe}.
Similarly, the rotating generalization of the Schwarzschild solution 
was found only in 1963 by Kerr\cite{Ker}.
The paradigm which developed may perhaps be marked by 
the 1973 article of Bardeen, Carter \& Hawking\cite{BCH}, 
enunciating the four laws of black-hole mechanics,
or the 1973 textbooks of Hawking \& Ellis\cite{HE}
and Misner, Thorne \& Wheeler\cite{MTW}.
Reviews in Hawking \& Israel\cite{HI} 
stylishly etch this theory in stone.
However, a few have come to the view that 
a more {\em local, dynamical} paradigm is required,
for instance to understand the black-hole collisions 
which are expected to be observed by gravitational-wave detectors.
The millenial Marcel Grossmann meeting provided a timely opportunity
to bring together some of the few (we happy few) 
whose research relates to this theme.

The standard paradigm for black holes 
consists mainly of statics and asymptotics.
By {\em statics} I mean the study of stationary black holes 
and perturbations thereof.
Here a black hole is defined by a Killing horizon, 
with which one can associate a surface area $A$ 
and a surface gravity $\kappa$.
(Throughout the article, definitions and notation are referred to cited texts,
in the spirit of a review).
Uniqueness theorems restrict the class of solutions,
in vacuo to the Kerr black holes 
parametrized by mass $m$ and angular momentum $J$. 
A first law relates perturbations by
\begin{equation}
\delta m={\kappa\delta A\over{8\pi}}+\Omega\delta J
\label{static}
\end{equation}
where $\Omega$ is the angular velocity of the horizon.
A zeroth law expresses the constancy of $\kappa$
and a third law excludes reducing $\kappa$ to zero, 
for instance by test-particle perturbations.
The terminology here is analogous to that of thermodynamics,
with which a genuine connection was conjectured by Bekenstein
and found by Hawking and others:
quantum fields radiate from stationary black holes 
with a black-body spectrum at temperature $\kappa/2\pi$.
Thus black holes presumably have an entropy $A/4$,
leading to much speculation concerning desirable quantum gravity.

With regard to this unanticipated connection,
it is interesting to note that 
thermodynamics itself is undergoing a radical paradigm shift 
from the theory promulgated in old-fashioned textbooks and lecture courses,
in which the first and second laws are formulated with derivatives 
which are either static state-space perturbations or proudly meaningless,
to a local, dynamical theory\cite{th}
in which the first and second laws are local field equations 
involving space-time derivatives of tensorial fields,
just as in the rest of physics.
This originated in 1940 
with the revolutionary work of Eckart\cite{Eck} on relativistic thermodynamics,
but the non-relativistic version apparently took decades longer\cite{Tru}
and is still foreign to most thermodynamics textbooks.
Such historical oddities and timescales clearly show that
science is less logical and more sociological than its preferred public image.
Will black-hole physics fare better?

By {\em asymptotics} I mean the study of conformal infinity 
in asymptotically flat space-times.
Penrose\cite{PR} showed how conformal transformations $g\mapsto\Omega^2g$
can be used to define conformal boundaries of the space-time, 
where $\Omega=0$, which are at infinite distance and/or time.
At future null (lightlike) infinity $\Im^+$,
the Bondi mass-energy $E$ measures the unradiated energy of the space-time,
while the Bondi flux $\varphi_-$ measures the energy flux $\psi=\Omega^2\varphi$
of the gravitational waves or other radiation.
They are related by the Bondi energy-loss equation
\begin{equation}
\nabla_-E=\oint\hat{*}\varphi_-
\label{Bondi}
\end{equation}
with the corresponding equation for past null infinity $\Im^-$ 
obtained by interchanging $\pm\to\mp$.
The original work of Bondi and others is acknowledged as the first 
to show conclusively that gravitational waves carry energy.

\begin{figure}
\centerline{\epsfxsize=6cm \epsfbox{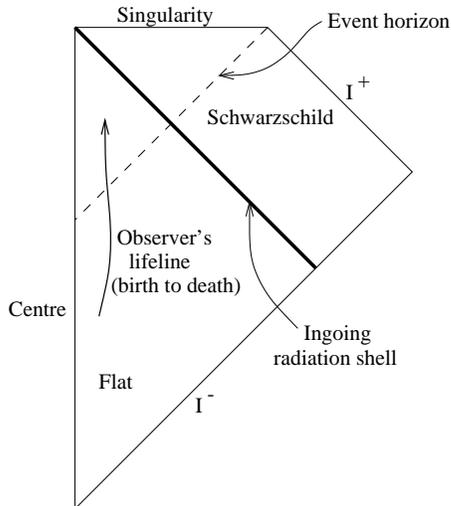}}
\caption{Example of the unknowable nature of event horizons:
Penrose diagram of a spherically symmetric space-time
which is initially flat but contains an ingoing radiation shell,
forming a Schwarzschild black hole.
An observer in the flat region crosses the event horizon but feels nothing.
The intrepid observer lives a full, happy and productive life,
but passes away before any possible knowledge of curved space-time.}
\label{event}
\end{figure}

The wrong turn occured, in my view, 
by applying asymptotic concepts to define black holes.
Existing textbooks define a black hole by an event horizon,
a phrase popular enough to title a recent film.
This is the boundary of the causal past of $\Im^+$,
meaning that it is defined by a boundary condition 
applied infinitely far in the future, about which we can know nothing,
according to relativistic causality.
The location of the event horizon, or even its existence, 
is known only after the universe has ended, 
or, depending on one's religious beliefs, 
to the gods looking down on space-time as a vast Penrose diagram.
It cannot be known to mere mortals in the here and now.
This is illustrated in Fig.\ref{event}:\footnote
{For the uninitiated, a Penrose diagram compactly summarizes 
the causal features of a space-time, as far as two dimensions allow.
Light rays run diagonally upwards,
separating spatial (sideways) directions 
from future (upwards) and past (downwards) directions.
Also, 
conformal boundaries at infinite distance and/or time are rendered finite.}
an observer crossing the event horizon
feels no gravitational field at all 
and has never experienced curved space-time.
The event horizon does not have any physical effect.
Such a horizon could be passing through you, gentle reader, 
at any given instant;
no-one would notice.
More realistically, 
since the actual universe is not thought to be asymptotically flat,
event horizons do not actually exist,
and therefore neither do black holes, by the standard definition.
This may come as some surprise to those observing black holes 
or their signatures;
I mention it only to highlight the shameful state of accepted black-hole theory.

More practical researchers, such as those making numerical simulations, 
usually characterize black holes by {\em marginal surfaces},
spatial surfaces which are extremal in a null hypersurface.
Marginal surfaces are often described by the derogatory term apparent horizon, 
which leads to confusion with the textbook definition of the latter,
due to Hawking\cite{HE}.
Apparent horizons are defined in asymptotically flat space-times:
one chooses an asymptotically flat spatial hypersurface
and finds all outer trapped surfaces lying in the hypersurface;
the boundary of this outer trapped region is the apparent horizon 
of the hypersurface.
However, this is so slicing dependent that 
there are global slicings of the Schwarzschild black hole 
with no apparent horizon\cite{WI}.
Moreover, even with a well chosen slicing of a general space-time,
it is impractical to check every embedded surface 
to see whether it is outer trapped.
In practice, people use marginal surfaces,
based on the proposition of Hawking\cite{HE} that 
a suitably smooth apparent horizon is a marginal surface, cf.\cite{KH}.
Algorithms exist to find marginal surfaces in numerical simulations,
as described by Deirdre Shoemaker during the session.
For instance, a black-hole coalescence may be defined by the appearance of 
a family of marginal surfaces enclosing the two original families.

Marginal surfaces seem, then, to provide the way forward.
However, 
without further qualification they cannot be taken to define black holes,
as they can occur in white-hole horizons, Cauchy horizons, 
cosmological horizons and wormhole horizons,
and do occur through every point of de~Sitter space-time.
My own proposal\cite{bhd,bhs} for the required refinement, 
{\em trapping horizons}, 
is described in the following.
An earlier suggestion by Tipler\cite{Tip} 
was compared by Brien Nolan during the session.
More recently, 
Ashtekar\cite{ABF,ABK} has proposed using {\em isolated horizons},
which are essentially null trapping horizons.
A significant body of work is currently being developed on isolated horizons,
as reported by Jerzy Lewandowski, Olaf Dreyer and Alejandro Corichi.
Similar ideas for cosmological horizons were presented by Jun-ichiro Koga,
and Daisuke Ida discussed both trapping horizons and apparent horizons.
The session also included various work on conservation laws,
Killing horizons and event horizons.

\section{Trapping horizons}

Imagine enclosing a star with a roughly spherical spatial surface 
at some moment of time.
Imagine detonating a flash of light simultaneously at each point of the surface.
Two wavefronts form, one ingoing and one outgoing.
Normally one expects the outgoing wavefront to have increasing area
and the ingoing wavefront to have decreasing area.
This is measured at each point by the expansions $\theta_\pm$ 
of light rays in the wavefronts:
$\theta_+>0$ for the outgoing wavefront 
and $\theta_-<0$ for the ingoing wavefront.
However, the gravitational field of the star tends to drag things toward it,
including light.
Thus the outgoing wavefront does not expand as much as if 
the star were not present.
The effect increases closer to the star and for larger mass.
For large enough mass, for a surface close enough,
it may happen that the outgoing wavefront has decreasing area.
This is a black hole:
all light and therefore matter are confined inside a shrinking area.
The idea then is that 
the boundary of the black hole at a given time is a marginal surface,
meaning that one wavefront has instantaneously parallel light rays,
in this case $\theta_+=0$.
To characterize a black hole, it is also important that 
the ingoing wavefront is converging, $\theta_-<0$,
and that $\theta_+$ is decreasing in the ingoing direction.

With such reasoning, I defined a {\em trapping horizon}\cite{bhd} 
as basically a hypersurface foliated by marginal surfaces.
As above, a marginal surface is a spatial surface, usually assumed compact,
on which one null expansion vanishes, fixed here as $\theta_+=0$.
I call the horizon {\em future} or {\em past}
if $\theta_-<0$ or $\theta_->0$ respectively,
and {\em outer} or {\em inner} 
if $\partial_-\theta_+<0$ or $\partial_-\theta_+>0$ respectively.
Here the null derivatives $\partial_\pm$ are with respect to 
a double-null foliation adapted to the horizon,
i.e.\ two intersecting families of null hypersurfaces 
whose intersections include the marginal surfaces.
Then I propose defining a non-degenerate {\em black hole} 
by a future outer trapping horizon.\footnote
{Degenerate black holes are those for which 
$\theta_+$ decreases in the $\partial_-$ direction,
but $\partial_-\theta_+$ is not strictly negative, 
for instance the extreme (maximally charged) Reissner-Nordstr\"om black hole.
However, they are not expected to be physically attainable;  
indeed, this is one possible formulation of the third law, 
as subsequent definitions of surface gravity reveal.}
More precisely, 
I suggest that a non-degenerate black hole exists only if such a horizon exists.
As to the converse, I do not wish to rule out strengthening the definition.
For instance, it seems reasonable to expect 
any surface sufficiently close to a marginal surface and inside the horizon 
to be a trapped surface, meaning $\theta_+\theta_->0$.
In contrast, the definition cannot be weakened 
without losing some basic properties expected of black holes,
as described in the next section.

\begin{figure}
\centerline{\epsfxsize=6cm \epsfbox{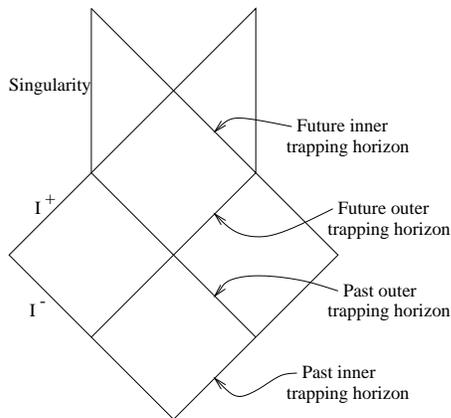}}
\caption{Penrose diagram of a generic Reissner-Nordstr\"om black hole,
an Einstein-Maxwell solution,
indicating the types of trapping horizon.
The diagram is left-right symmetric and identified at top and bottom.
In the uncharged (Schwarzschild) case, only the outer horizons exist.}
\label{rn}
\end{figure}

\begin{figure}
\centerline{\epsfxsize=7cm \epsfbox{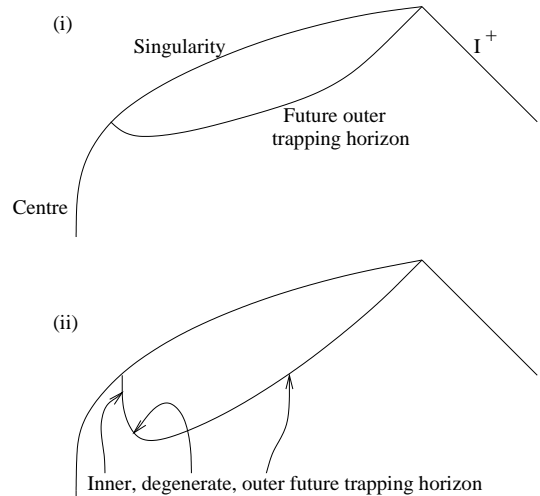}}
\caption{Penrose diagrams of gravitational collapse to a black hole
in spherical symmetry.
(i) depicts the generic collapse of a massless Klein-Gordon field,
where the trapping horizon is always of the outer type,
while (ii) illustrates that, for other matter models,
the horizon may have an inner part, separated by a degenerate part.
As shown in the next section, 
the horizon is achronal or causal if it is outer or inner respectively.
Thus in case (ii), the degenerate part occurs
where the trapping horizon is null.}
\label{collapse}
\end{figure}

The four non-degenerate types of trapping horizon occur for 
the generic Reissner-Nordstr\"om black hole, 
where they label the various Killing horizons as the terminology suggests,
as shown in Fig.\ref{rn}.
Examples of gravitational collapse to a black hole 
are depicted in Fig.\ref{collapse}.
For the massless Klein-Gordon field in spherical symmetry,
Christodoulou\cite{Chr} showed that 
generic collapse satisfies cosmic censorship,
in the sense that a spatial singularity forms inside a future trapping horizon.
In this case,
it is straightforward to show that a trapping horizon must be of the outer type.
However, 
other matter models such as dust allow the horizon to have an inner part.

\section{General laws}

I showed that several fundamental properties of trapping horizons
follow directly from the Einstein equations,
provided that the matter satisfies a standard local energy condition.
The dominant energy condition states that 
a future-causal observer measures future-causal or zero energy-momentum.
This implies the weak energy condition, 
which states that a causal observer measures non-negative energy density.
This in turn implies the null energy condition, 
which states that the energy density in a null frame is non-negative.

The {\em signature law}\cite{bhd} states that 
an outer (respectively inner) trapping horizon 
is achronal (respectively causal),
assuming the null energy condition.
Here achronal means spatial (spacelike) or null 
and causal means temporal (timelike) or null.
The null case is non-generic in the sense that 
special (local stationarity) conditions are required on the matter and geometry.
Thus a black-hole horizon is generically spatial and always one-way traversible:
one can enter but not leave the horizon.
This is a fundamental property expected of black holes,
without which a definition would not be acceptable.

The {\em second law}\cite{bhd} states that a future outer or past inner 
(respectively past outer or future inner) trapping horizon 
has non-decreasing (respectively non-increasing) area element,
again assuming the null energy condition.
Here the orientation is such that, in the null limit, it is future-null.
Thus for a black-hole horizon,
\begin{equation}
A'\ge0
\label{second}
\end{equation}
where $A$ is the area of the marginal surfaces
and the prime denotes the derivative 
along a vector generating the marginal surfaces, with the above orientation.
Moreover, $A'$ vanishes if and only if the horizon is null,
which includes both Killing horizons and isolated horizons.
Thus the area of a black hole cannot decrease and generically increases.
Again this is an expected property of black holes,
analogous to the second law of thermodynamics.
Note the difference with the textbook second law for event horizons,
due to Hawking, who conjectured such a property of apparent horizons\cite{HE}.

The {\em topology law}\cite{bhd} states that 
future or past outer trapping horizons have 
topologically spherical marginal surfaces,
assuming the dominant energy condition.
Again this is an expected property of black holes.
If one allows degenerate black holes, then toroidal topology is possible,
but highly non-generic, e.g.\ the marginal surface must be Ricci flat.
Similarly, 
if one allows non-compact marginal surfaces, planar topology is possible,
but this is hardly of astrophysical relevance.
Non-orientable topology is ruled out by the future or past condition.

Another general result obtained by these methods is that,
in the presence of a positive cosmological constant $\Lambda$,
black holes have an {\em area limit}\cite{HSN}: outer trapping horizons satisfy
$A\le4\pi/\Lambda$.
Thus black holes are smaller than the cosmological horizon scale,
corresponding to an area $12\pi/\Lambda$.

In summary, trapping horizons provide a local, dynamical definition 
of a black hole with expected properties.
The definition is practical in the sense that 
marginal surfaces can be found in numerical simulations,
with the future outer condition being straightforward to check.
It is also simple and intuitive:
the horizon is where outgoing light rays are just trapped 
by the gravitational field.
The next phase of research consists of 
identifying the relevant physical quantities
and finding the equations relating them.
This should include and generalize what is known in statics and asymptotics,
principally the first law (\ref{static})
and the Bondi energy-loss equation (\ref{Bondi}) respectively.
Thus we need local, dynamical definitions of quantities such as 
mass-energy $E$ and $m$, surface gravity $\kappa$, 
energy flux $\psi$ of gravitational waves and so on.
This constitutes the framework I call {\em black-hole dynamics}.
So far, the program has been completed only in symmetric cases,
discussed in the next two sections.

Without the local energy conditions,
trapping horizons may also be used to define 
dynamic (traversible) wormholes\cite{wh}.
Then, depending on the sign of the energy density of the matter,
black holes and wormholes are locally interconvertible.
For example, an initially Schwarzschild black hole evaporates 
by Hawking radiation as a dynamic wormhole.
This leads to a theory of {\em wormhole dynamics}, to be described elsewhere.

\section{Spherical symmetry}

The black-hole dynamics framework has been developed in detail 
in spherical symmetry\cite{1st}. 
Here the area $A$ of the spheres is a geometrical invariant.
Since the null expansions are $\theta_\pm=\partial_\pm\ln A$,
the basic definitions can be formulated and extended in terms of $A$,
or more conveniently, the areal radius $r=\sqrt{A/4\pi}$:
a sphere is untrapped, marginal or trapped 
as $\nabla r$ is spatial, null or temporal respectively.
A trapping horizon is outer, degenerate or inner
as $\nabla^2r$ is positive, zero or negative respectively.

There is a natural choice of time given by the Kodama vector $k$,
defined up to orientation by $k\cdot\nabla r=0$
and $k\cdot k=-\nabla r\cdot\nabla r$.
This reduces to the usual Killing vector 
for Schwarzschild or Reissner-Nordstr\"om black holes.
In general, a sphere is trapped, marginal or untrapped
if $k$ is spatial, null or temporal respectively.
Thus a trapping horizon occurs where $k$ is null,
just as a Killing horizon occurs where the Killing vector is null.
A noteworthy difference is that, for Killing horizons,
the normalization of the Killing vector is determined by 
an asymptotic boundary condition,
that it be unit at infinity,
whereas the normalization of $k$ is fixed locally.

The {\em energy-momentum density} with respect to $k$ is the vector
$j=-T\cdot k$,
where $T$ is the energy-momentum-stress tensor of the matter.
Then both $k$ and $j$ are covariantly conserved:
\begin{eqnarray}
&&\nabla\cdot k=0\\
&&\nabla\cdot j=0.
\end{eqnarray}
These Noether currents therefore admit Noether charges
\begin{eqnarray}
&&V=-\int_\Sigma\hat{*}\cdot k=\textstyle{4\over3}\pi r^3\\
&&E=-\int_\Sigma\hat{*}\cdot j=\textstyle{1\over2}r(1-\nabla r\cdot\nabla r)
\end{eqnarray}
defined on each sphere,
i.e.\ independent of the choice of spatial hypersurface $\Sigma$ 
with regular centre.
Then $V$ is the areal volume, while $E$ is the desired {\em energy}.
This definition of mass or energy is common 
and appears to have been given first by Misner \& Sharp.
It has many desired physical properties\cite{sph},
including that it reduces to the Bondi energy at $\Im^\pm$.

I define {\em surface gravity} $\kappa$ by
\begin{equation}
k\cdot(\nabla\wedge k)=\kappa\nabla r.
\end{equation}
That is, the vectors on each side of the equation are parallel,
and $\kappa$ is defined as the proportionality factor.
On a trapping horizon $\partial_\pm r=0$, 
$k=\pm\nabla r$, so that the equation takes the same form
as the usual definition of stationary surface gravity 
in terms of the stationary Killing vector.
Thus we have a natural generalization of surface gravity 
for dynamic black holes.
Note as above that the normalization of $\kappa$ is determined locally,
without recourse to asymptotic boundary conditions.
The surface gravity evaluates as $\kappa=\nabla^2 r/2$ on a trapping horizon,
and therefore has the desired property that
$\kappa>0$, $\kappa=0$ or $\kappa<0$ 
on outer, degenerate or inner trapping horizons respectively.
Thus any formulation of the {\em third law} expressing $\kappa\not\to0$
is related to the unattainability of degenerate black holes.

Associated with the matter are two invariants of $T$,
an energy density $w$, which is interpreted as a {\em work density},
and an {\em energy flux} covector $\psi$.
The dominant energy condition implies $w\ge0$
and the null energy condition implies that 
$\psi$ is outward achronal in untrapped regions.
At null infinity $\Im^\pm$, 
projecting $r^2\psi$ along $\partial_\mp$ 
yields the Bondi flux $\varphi_\mp$.
Thus the Bondi flux has been localized for any sphere in the space-time.

The Einstein equation implies what I call the {\em unified first law}
\begin{equation}
\nabla E=A\psi+w\nabla V.
\label{unified}
\end{equation}
The two terms are interpreted as energy-supply and work terms respectively, 
analogous to the heat supply and work of the first law of thermodynamics.
Projecting the equation along $\Im^+$, 
it reduces to the Bondi energy-loss equation (\ref{Bondi}),
since $A=\oint\hat{*}r^2$ and $w\to0$.
Generally
\begin{equation}
\nabla_\pm E=A\psi_\pm
\end{equation}
at $\Im^\mp$.
Thus the Bondi energy loss has been localized.

Projecting the unified first law (\ref{unified}) along a trapping horizon,
\begin{equation}
E'={\kappa A'\over{8\pi}}+wV'
\label{first}
\end{equation}
where the prime denotes the same derivative as in the second law (\ref{second}).
The term involving area and surface gravity has the same form as that of
the textbook first law (\ref{static}) of black-hole statics, 
with a space-time derivative replacing the perturbation.
Thus I call the equation the {\em first law of black-hole dynamics}.
The work term may be checked for the Maxwell field,
for which $w={\cal E}^2/8\pi$ is the energy density 
of an electric field ${\cal E}=e/r^2$, where $e$ is the charge.
Note that for the Reissner-Nordstr\"om black hole,
the energy is $E=m-e^2/2r$, 
so that the form of the work term depends on whether 
$E$ or $m$ appears on one side of the first law.
In general, one knows $E$ but not $m$, 
since there is no known local prescription for the latter 
which works for all matter fields.

Thus the two principal equations of statics and asymptotics,
the first law (\ref{static}) and the Bondi energy-loss equation (\ref{Bondi}),
have been unified into a single energy conservation equation (\ref{unified})
which holds throughout the space-time.
This unified first law also includes a 
{\em first law of relativistic thermodynamics}:
projecting it along the flow of a thermodynamic material yields
\begin{equation}
\dot E=\alpha\dot Q-p\dot V
\end{equation}
where the dot denotes the material derivative,
$p$ is the radial pressure, $\dot Q$ is the heat supply
and $\alpha$ is a red-shift factor.
Apart from this Tolman-like factor, which is 1 in the Newtonian limit,
the equation has the same form as the textbook first law of thermodynamics.
Thus a genuine connection with thermodynamics has been found,
to be added to the famous connection between surface gravity and temperature.
This also suggests that 
even dynamic black holes have an entropy $A/4$\cite{HMA}.

The Einstein equation yields an explicit expression for the surface gravity:
\begin{equation}
\kappa={E\over{r^2}}-4\pi rw.
\end{equation}
Apart from the matter term,
this has the same form as Newton's law of gravitation.
This allows various inequalities 
relating area, energy and surface gravity\cite{in}.
It also implies a {\em zeroth law}, given here for the first time:
if $w'=0$ along a null trapping horizon, then
\begin{equation}
\kappa'=0.
\label{zeroth}
\end{equation}
The null condition is an expression of 
local stationarity of the trapping horizon, $A'=0$, 
while $w'=0$ is a corresponding expression for the matter, 
e.g.\ ${\cal E}'=0$ or $e'=0$ for an electric field.
The space-time need not be stationary, 
as emphasized for isolated horizons\cite{ABF},
so this is a non-trivial generalization of the textbook zeroth law.

In summary, 
the black-hole dynamics framework is essentially complete in spherical symmetry:
one knows the relevant physical quantities $(A,k,E,\kappa,\psi,w)$
and the equations relating them.
There is a unified first law and, for black holes,
zeroth, first and second laws, which all involve the same derivative,
that generating the trapping horizon.
Compare here with the textbook zeroth, first and second laws, 
which all involve different derivatives, 
except where the second law reduces to an equality.
This is ironically reminiscent of the confusion of derivatives 
which plagued thermodynamics; 
see Truesdell\cite{Tru} for a critical history.

\section{Cylindrical symmetry}

The black-hole dynamics framework has also been applied 
in cylindrical symmetry\cite{cyl},
which has the additional complexity of gravitational waves.
Here the basic geometrical invariants are the circumferential radius $\rho$
and, up to a constant scale, the specific length $\ell$ of the cylinders.
Writing $r=\rho\ell$, 
the definitions of trapped, marginal and untrapped surfaces, 
and outer, degenerate and inner trapping horizons,
all take the same form as in spherical symmetry.
There is a canonical time vector $k$ defined by the same formulas,
which is covariantly conserved,
$\nabla\cdot k=0$,
with Noether charge $V=\pi r^2$.
An important difference with spherical symmetry is that 
the corresponding energy-momentum density per specific length of the matter,
$j[T]=-\ell^{-2}T\cdot k$,
is generally not conserved.
Physically this is because gravitational waves carry energy.

Remarkably, it is possible to include the energy of the gravitational waves 
in a combined conservation law.
The key physical quantity is the {\em gravitational potential} $\phi=-\ln\ell$.
One invariant of the Einstein equation is a wave equation for $\phi$:
\begin{equation}
\nabla^2\phi=4\pi\varrho
\label{poisson}
\end{equation}
where $\varrho$ is an invariant of $T$.
In the Newtonian limit, 
this reduces to the Poisson equation of Newtonian gravity,
with $\varrho$ reducing to the density
and $\phi$ reducing to the Newtonian gravitational potential.
Then I define the 
{\em energy-momentum-stress tensor of the gravitational waves} 
\begin{equation}
\Theta={2\nabla\phi\otimes\nabla\phi-(\nabla\phi\cdot\nabla\phi)g\over{8\pi}}
\end{equation}
which has the Klein-Gordon form.
Then the energy-momentum density per specific length 
of the gravitational waves is
$j[\Theta]=-\ell^{-2}\Theta\cdot k$.
The combined energy-momentum density is then covariantly conserved:
\begin{equation}
\nabla\cdot j[T+\Theta]=0.
\end{equation}
The corresponding Noether charge is an energy per specific length,
$E={1\over8}(1-\ell^{-2}\nabla r\cdot\nabla r)$,
originally due to Thorne.

With the energy of the gravitational waves included,
definitions and results follow analogously to those in spherical symmetry.
Surface gravity $\kappa$ may be defined by
$k\cdot(\nabla\wedge k)=\ell\kappa\nabla r$.
The matter admits a work density $w$ and an energy flux $\psi[T]$,
with a corresponding $\psi[\Theta]$, 
the {\em energy flux of the gravitational waves}.
(The original reference\cite{cyl} used $\psi/\ell$).
Then the Einstein equation implies the unified first law
\begin{equation}
\ell\nabla E=A\psi[T+\Theta]+\ell^{-1}w\nabla V
\end{equation}
where $A=2\pi r$ is the specific area.
The first law of black-hole dynamics is the projection along a trapping horizon:
\begin{equation}
\ell E'={\kappa A'\over{8\pi}}+{wV'\over\ell}.
\end{equation}
The zeroth law is $\kappa'=0$
along a null trapping horizon with $(w/\ell)'=0$.
A first law for cosmic strings also follows 
by projecting the unified first law along the string.

\section{Current directions}

Of the physically important quantities familiar from statics and asymptotics,
those not yet considered are angular momentum and angular velocity.
For instance, in a black hole collision, 
one expects that a certain proportion of the initial angular momentum 
will be radiated away as gravitational waves.
Thus angular momentum should enter the first law, 
as in the static first law (\ref{static}),
and presumably satisfy a conservation law of its own, as in Newtonian physics.
However, there is no agreed local definition of angular momentum 
for dynamic black holes, even in symmetric cases.
The natural arena seems to be axisymmetry, 
though twisted cylindrical symmetry might prove simpler.
New ideas are needed here.
One idea is a Noether-current method\cite{noe} 
which recovers the Komar integrals 
for both energy and angular momentum in vacuo,
but allows generalization to certain matter fields. 

Also needed are generalizations of the physical quantities 
already defined in spherical or cylindrical symmetry.
For instance, 
Mukohyama\cite{MH} proposed a general definition of surface gravity
\begin{equation}
\kappa={1\over{16\pi r}}
\oint{*}g^{+-}(2\partial_{(+}\theta_{-)}+\theta_+\theta_-)
\end{equation}
where $r=\sqrt{A/4\pi}$, $A=\oint{*}1$ and the integrals are over 
spatial surfaces in the double-null foliation adapted to the trapping horizon.
This is a quasi-local rather than local definition,
as it requires knowledge of the whole surface.
The simplest definition of quasi-local energy 
generalizing the spherically symmetric energy, 
satisfying some similar properties\cite{mon}, is the Hawking energy
\begin{equation}
E={r\over{16\pi}}\oint{*}(R-g^{+-}\theta_+\theta_-)
\end{equation}
where $R$ is the Ricci curvature scalar of the surface.
Then there is a first law with the same form as (\ref{first}),
for a certain definition of $w$\cite{MH}, which now may be non-zero in vacuo.
This should presumably include angular momentum, 
but as above, this is not properly understood.
If this form of the first law is accepted,
there is a corresponding zeroth law 
just as in spherical symmetry (\ref{zeroth}).
As to a third law for dynamic black holes, 
this seems to be still a completely open question.

If general black holes prove to be beyond human understanding,
approximation methods may still be useful.
For instance, for linearized gravitational waves, 
it is well known how to construct an effective energy tensor\cite{MTW} 
analogous to the $\Theta$ of cylindrical symmetry.
I recently proposed a quasi-spherical approximation\cite{qs} 
which can be used to describe gravitational waves 
from roughly spherical black holes.
The physical quantities and laws of black-hole dynamics,
along with {\em gravitational-wave dynamics},
generalizing the relativistic Poisson equation (\ref{poisson}),
can be formulated in this context, to be described elsewhere.

\begin{figure}
\centerline{\epsfxsize=6cm \epsfbox{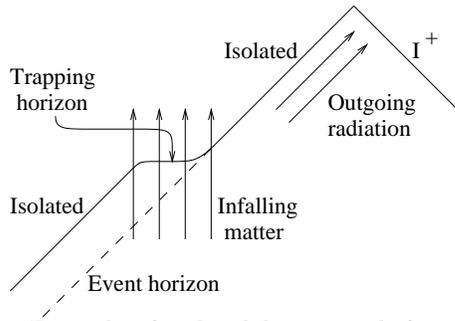}}
\caption{Example of isolated horizons:
before and after matter enters a black hole, 
the trapping horizon may be an isolated horizon.
The space-time need not be stationary;
there can be outgoing radiation arbitrarily close to the horizon.
The locally irrelevant event horizon is depicted for comparison.}
\label{isolated}
\end{figure}

Recently Ashtekar has encouraged work on isolated horizons,
which are null trapping horizons with certain additional conditions\cite{ABF}.
This is more general than traditional statics in that 
an isolated horizon is not necessarily a Killing horizon.
The situation can be thought of as analogous to dynamic equilibrium 
rather than static equilibrium:
the space-time need not be stationary,
but the part of interest, the black-hole horizon, is unchanging.
This can be used to describe black holes in quiescent states 
between dynamic processes, as depicted in Fig.\ref{isolated}.
First and zeroth laws for isolated horizons have been given\cite{ABF},
the second law reducing to an equality.
A particularly remarkable result is 
a quantum-geometrical derivation of black-hole entropy\cite{ABK}.

\section{Perspective}

This has been a personal view of black-hole theory based on current knowledge,
necessarily diverging from generally accepted theory.
I have presented a framework for investigating black-hole dynamics,
without pretence to a complete theory.
There is already ample evidence that this quest is successfully proceeding:
symmetric cases have been completely analyzed, 
with all physically important quantities known,
and there are general laws such as the second law,
with at least suggestions for general first and zeroth laws.
Further development requires further ideas,
which presents exciting opportunities for original research.
This is particularly timely given 
the anticipated era of gravitational-wave astronomy,
which promises unprecedented interplay between 
black-hole theory and observation.
I hope that the dawn of the new millenium will herald 
a shift towards a more local, dynamical understanding of black holes.

\bigskip\noindent
I am grateful to the organizers of the Ninth Marcel Grossmann Meeting 
for the invitation to organize the Generalized Horizons session, 
to the National Science Foundation for a travel grant,
and to the other contributors.

\end{document}